# Magnetism of σ-phase Fe-Mo alloys: ac magnetic susceptibility study

J. Przewoźnik and S. M. Dubiel[*]

AGH University of Science and Technology, Faculty of Physics and Applied Computer Science, al. A. Mickiewicza 30, PL-30-059 Kraków, Poland

A series of four σ-$Fe_{100-x}Mo_x$ samples was investigated with ac magnetic susceptibility measurements. An evidence was found that the ground magnetic state of the samples is constituted by a spin glass (SG) with the spin glass temperature ranging between ~34K for $x$=47 and ~16K for $x$=53. The SG state is heterogeneous and it can be divided into a weak-irreversibility and a strong-irreversibility domains. Its figures of merit are typical of metallic (canonical) spin glasses.

[*]Corresponding author: Stanislaw.Dubiel@fis.agh.edu.pl

The sigma ($\sigma$) phase (space group $D^{14}_{4h}$ - $P4_2/mnm$ ) is one of many members of the Frank-Kasper (FK) phases [1]. It can be formed in alloys in which at least one constituting element is a transition metal. A tetragonal unit cell of $\sigma$ hosts 30 atoms distributed over 5 different lattice sites, usually termed as A, B, C, D and E. The sites have high coordination numbers (12-16), and the distribution of atoms is not stoichiometric. These features, in combination with the fact that $\sigma$ can be formed in a certain range of composition, make it that $\sigma$-phase alloys show a diversity of physical properties that can be tailored by changing constituting elements and/or their relative concentration. Structural complexity and chemical disorder make them also an attractive yet challenging subject for investigation. The interest in $\sigma$ is further justified by a deteriorating effect on useful properties of technologically important materials in which it has precipitated e. g. [2,3]. It should be, however, noticed that attempts have been undertaken to take advantage of its high hardness in order to increase materials strength and durability e. g. [4,5].

Concerning magnetic properties of $\sigma$ in binary alloys, until recently only $\sigma$ in Fe-Cr and Fe-V, systems was known to be magnetic and regarded as ferromagnetic [6-8]. Its magnetism has been recently revealed to be more complex than initially anticipated viz. its re-entrant character with a spin glass as the ground state was evidenced [9]. Furthermore, nuclear magnetic resonance measurements performed on the $\sigma$-FeV samples revealed that not only iron but also vanadium atoms present on all five sub lattices were magnetic [10]. Freshly, the magnetism of $\sigma$ was discovered in Fe-Re [11] and in Fe-Mo [12] systems.

Based on a systematic ac magnetic susceptibility measurements carried out on a series of four $\sigma$-$Fe_{100-x}Mo_x$ ($x$=45, 47, 51, 53) samples a clear evidence is here reported that it is the SG that constitutes the ground magnetic state of the $\sigma$-FeMo alloys. Its characteristic features are typical of metallic SGs i.e. such in which Ruderman-Kittel-Kasuya-Yosida (RKKY) interaction is active.

The samples of $\sigma$ were prepared by the following way: powders of elemental iron (3N+ purity)) and molybdenum (4N purity) were mixed together in appropriate proportions and masses (2g), and next pressed to pallets. The pellets were subsequently isothermally annealed at 1703 K during 6 h, and afterward quenched into liquid nitrogen. The mass loses of the fabricated samples were less than 0.01% of their initial values, so it is reasonable to assume that their real compositions are close to nominal ones. X-ray diffraction patterns recorded on the powdered samples gave evidence that their crystallographic structure was to more than 97.5 % $\sigma$. More details can be found elsewhere [13].

The ac magnetic susceptibility data were collected using the Quantum Design physical property measurement system (PPMS) between 2.0 and 100 K in a 2 Oe ac magnetic field (and zero external dc magnetic field) for frequencies varying from 10 Hz to 10 kHz. The data were collected during the cooling down and warming up runs.

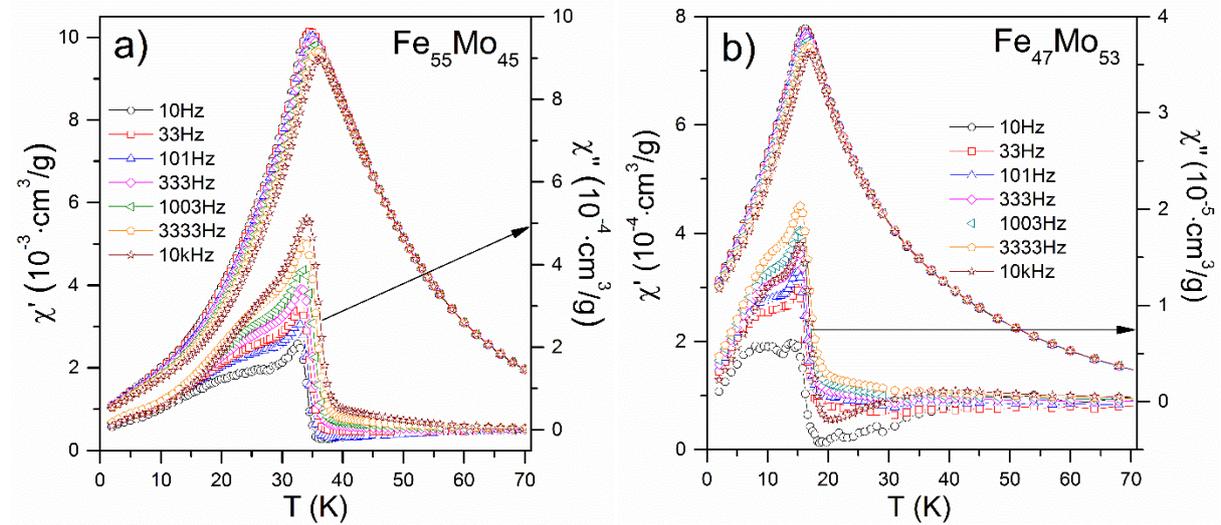

Fig. 1 (color online) Real ($\chi$') and imaginary ($\chi$'') parts of ac magnetic susceptibilities recorded vs. temperature, $T$, and various frequencies, $f$, shown on heating (up) for (a) the σ-$Fe_{55}Mo_{45}$ and (b) σ-$Fe_{47}Mo_{53}$ alloys.

The ac magnetic susceptibility curves, and in particular the $\chi$'($T$) ones, give a clear evidence viz. (1) a well-defined cusp, whose position, $T_f$, defines a spin-freezing temperature, and (2) its frequency shift towards higher temperature (Fig. 2a) that the studied samples exhibit a spin glass (SG) state. The $\chi$''($T$)-curves, despite lower intensities, show two characteristic temperatures: a higher one defined by an inflection point of $\chi$''($T$)-curve) which we will call $T_{if}$, and a lower one (ill-defined) which we will call $T_{si}$. The former temperature can be associated with the upper limit of a strong-irreversibility state of SG. As illustrated in Fig. 1a and in Fig. 1b, $T_{if}$ also shifts with $f$. The values of both $T_f$ and $T_{if}$ calculated for $f$=10 Hz are displayed in Table 1.

To further characterize the SG state, several quantities pertinent to SG have been calculated and analyzed below.

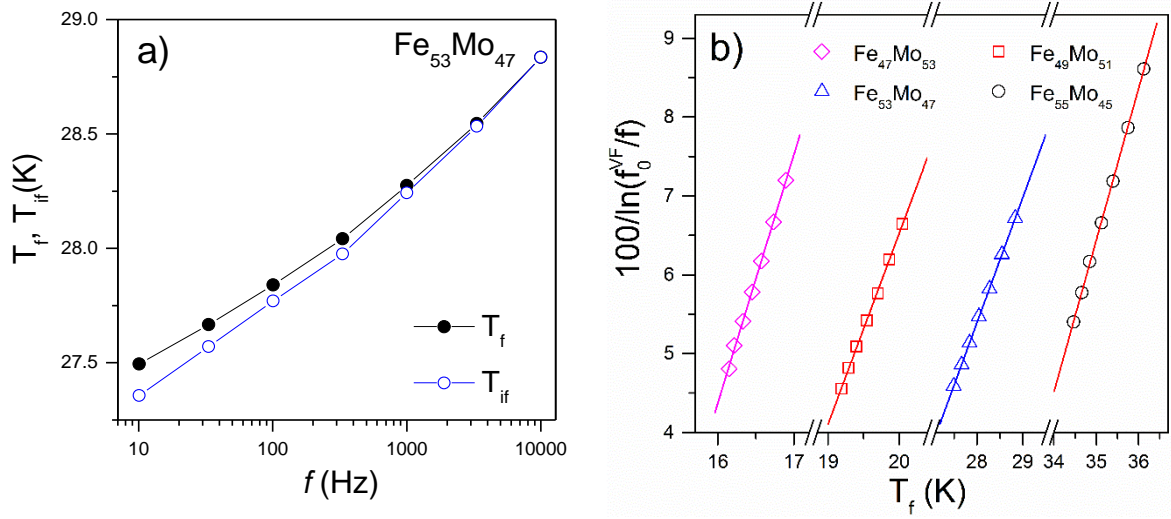

Fig. 2 (color online) (a) Spin freezing temperature vs. frequency, $f$, as found from $\chi'(T)$, $T_f$, and $\chi''(T)$, $T_{if}$, curves for σ-Fe$_{53}$Mo$_{47}$ alloy. The lines are drawn to guide the eye; (b) relationship between $T_f$ and frequency $f$ in terms of the Vogel-Fulcher law for σ-Fe$_{100-x}$Mo$_x$ alloys. The lines are the best fits to the data.

First, we calculated a relative shift of $T_f$ per a decade of frequency, $RST$, using the following equation:

$$RST = \frac{\Delta T_f / T_f}{\Delta \log f} \qquad (1)$$

The $RST$-values calculated both for $T_f$ and $T_{if}$ are shown in Table 1. They give a clear evidence that the frequency dependence of $T_f$ is weak, a feature characteristic of metallic (canonical) spin glasses in which the RKKY interaction is responsible for the magnetic coupling mechanism [12]. This is quite unexpected behavior as the concentration of magnetic carriers (Fe atoms) is very high in the studied samples.

Table 1 Best-fit values of characteristic quantities as obtained from a frequency ($f$) dependence of the cusp in the real ($\chi'$) and imaginary ($\chi''$) parts of the ac susceptibility. The meaning of the parameters is given in the text.

| $x$ | $T_f$ [K] | $T_{if}$ [K] | $T_o^{VF}$ [K] | $T_{SG}$ [K] | $RST_{\chi'}$ | $RST_{\chi''}$ | $E_a$ [K] | $f_0^{VF}$ [Hz] | $zv$ | $f_o$ [Hz] | $T_{si}$ [K] |
|---|---|---|---|---|---|---|---|---|---|---|---|
| 45 | 34.5 | 34.1 | 31.65 | 33.3 | 0.0128 | 0.0152 | 52(6) | $10^{9.0(4)}$ | 8.2(4) | $10^{13.1(4)}$ | 27 |
| 47 | 27.5 | 27.3 | 24.6 | 26.2 | 0.0135 | 0.0156 | 63(5) | $10^{10.5(3)}$ | 10.4(4) | $10^{14.8(4)}$ | 20.5 |
| 51 | 19.2 | 19.15 | 17.3 | 18.4 | 0.0122 | 0.0187 | 41(9) | $10^{10.5(9)}$ | 10(1) | $10^{15.0(9)}$ | 14 |
| 53 | 16.15 | 16.15 | 15.5 | 15.5 | 0.0128 | 0.0145 | 31.7(11) | $10^{10.0(1.4)}$ | 9.5(2) | $10^{14.3(1.5)}$ | 11.5 |

The frequency dependence of $T_f$ can be either analyzed in terms of an Arrhenius law i.e. the following equation:

$$f = f_o \exp(-\frac{E_a}{k_B T_f}) \qquad (2)$$

or in terms of a Vogel-Fulcher law i.e. the following equation:

$$f = f_0^{VF} \exp(-\frac{E_a}{k_B(T_f - T_0^{VF})}) \qquad (3)$$

where $E_a$ is an anisotropy (activation) energy, $k_B$ is the Boltzmann constant, and $T_0^{VF}$ is the Vogel-Fulcher temperature interpreted as the one at which the largest time scale for cluster relaxation diverges [13].

Attempts to fit the frequency dependence of $T_f$, with eq. (2), although mathematically successful, resulted in unphysical values of the frequency factor $f_o$ and of $E_a$. The former had values in the range of $10^{66}$-$10^{71}$ Hz, while the latter in the range of 2475-5126 K/$k_B$. On the other hand, equation (3) yielded reasonable values of all free parameters i.e. $f_o$, $E_a$ and $T_o$, as displayed in Table 1. Their values are similar to the corresponding ones found for the Cu-Mn spin glass regarded as the canonical one [14]. Finally, we have analyzed the frequency dependence of $T_f$ in terms of the critical slowing-down dynamics govern by the following equation:

$$f = f_o(\frac{T_f}{T_{SG}} - 1)^{zv} \qquad (4)$$

where $T_{SG}$ is a spin-glass temperature and the $zv$ is known as a dynamic exponent [15]. The data for all studied samples could have been successfully analyzed in terms of eq. (4), and the plots obtained are shown in Fig. 3.

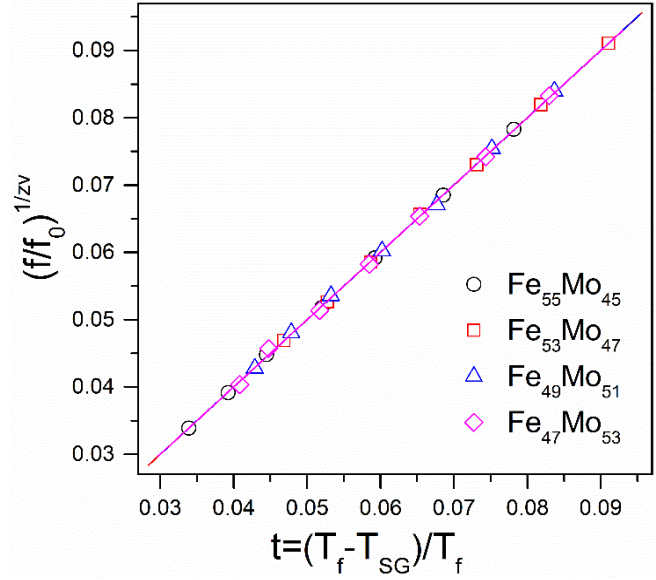

Fig. 3 (Color online) Relationship between frequency, $f$, and spin freezing temperature, $T_f$. The solid line is the best fit to the data presented in terms of the equation (4).

The values of the fit parameters involved in eq. (4) are included in Table 1. It can be seen that the values of $T_{SG}$ agree well with $T_f$ ones, those of $zv$ are in the range of 8-10 which is characteristic of canonical spin glasses [16] and $f_o$ vary between $10^{13}$ and $10^{15}$ Hz.

The lower temperature at which an anomaly in the $\chi''(T)$ curves can be seen, $T_{si}$, was determined as temperature at which the curve recorded for $f$=10 Hz has an inflection point. This temperature can be associated with the Gabay-Toulouse line which marks a transition from a weak into a strong irreversibility mode of SG.

Based on the above discussed results a magnetic phase diagram for the σ-FeMo alloys has been constructed and is shown in Fig. 4. Temperature characteristic of a magnetic transition obtained from Mössbauer measurements, $T_{MS}$, have been added [10]. A general agreement between the present ac and the former data [10] concerning the SG state can be seen. In particular, it is evident that SG is magnetically not homogenous, and it can be divided into two domains viz. (a) weak irreversibility (at higher temperatures) and (b) strong irreversibility (at lower temperatures) ones. The Mössbauer spectroscopic measurements revealed a magnetic transition at higher temperature than $T_f$ which means that the magnetism of the investigated samples may have a reentrant character. Positive values of the Curie-Weiss temperature, $\theta_{CW}$, derived from the temperature dependence of the inversed $\chi'$ susceptibility - displayed in Fig. 4 - indicate a ferromagnetic-like interactions that strongly depend on the alloy composition.

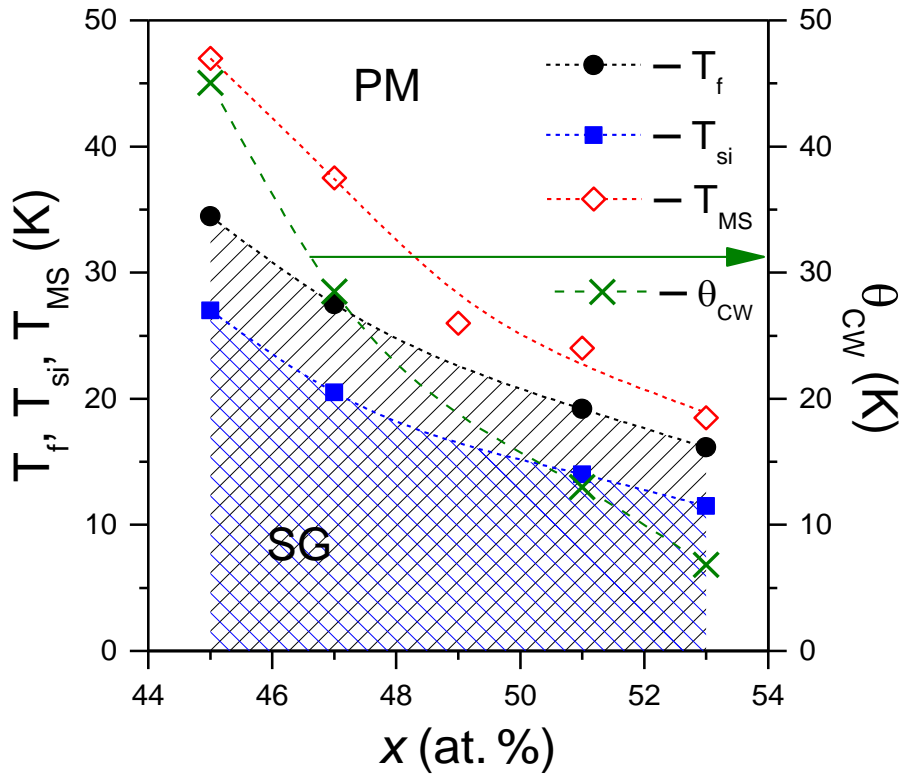

Fig. 4 (Color online) Tentative magnetic phase diagram of σ-$Fe_{100-x}Mo_x$ alloys. Full symbols and crosses represent the data obtained from the present ac susceptibility while open symbols indicate the data acquired from the Mössbauer measurements [10]. The lines are to guide the eye. The range of the SG regime is divided into the one with a strong irreversibility (diagonal crossing) and that with a weak irreversibility (diagonal-bottom-top).

To summarize, a magnetism in form of a spin glass was revealed as the ground magnetic state in the σ-phase Fe-Mo alloys. Its characteristic features viz. the relative shift of the spin freezing temperature per decade of frequency and the dynamic exponent are in line with those typical of the canonic spin glasses. An increase of the Mo concentration gradually decreases the spin glass temperature from ~34K at $x=45$ to ~16K at $x=53$. The present finding i.e. the existence of SG as the ground magnetic state is in agreement with the magnetism observed in other σ-phase binary alloys viz. Fe-Cr, Fe-V and Fe-Re.

**Acknowledgements**

This work was supported by The Ministry of Science and Higher Education of Polish government. J. Cieślak synthetized the samples.